\documentclass{appolbMOD}
\usepackage{graphicx}
\usepackage{mathptmx}
\usepackage{amsmath}
\usepackage{amssymb}
\usepackage{bm}

\newcommand{\pp}{\mathbf{p}}

\newcommand{\be}{\begin{equation}}
\newcommand{\ee}{\end{equation}}
\newcommand{\bea}{\begin{eqnarray}}
\newcommand{\eea}{\end{eqnarray}}
\newcommand{\ba}{\begin{align}}
\newcommand{\ea}{\end{align}}

\newcommand{\rme}{{\rm e}}

\usepackage{setspace}

\begin{document}
\title{Dense Matter and Neutron 
Stars: Some Basic Notions
}
\author{C. J. Pethick
\address{The Niels Bohr International Academy, The Niels Bohr Institute,\\ University of Copenhagen, Blegdamsvej 17, DK-2100 Copenhagen \O, Denmark}
\address{NORDITA, KTH Royal Institute of Technology and Stockholm University, Roslagstullsbacken 23, SE-106 91 Stockholm, Sweden}
}
\maketitle
\begin{abstract}
A number of properties of dense matter can be understood semiquantitatively in terms of simple physical arguments.  We begin with the outer parts of neutron stars, and consider the density at which pressure ionization occurs, the density at which electrons become relativistic, the density at which neutrons drip out of nuclei, and the size of the equilibrium nucleus in dense matter.  Subsequently, we treat the so-called ``pasta'' phases expected to occur at densities just below the density at which the transition from the crust to the liquid core of a neutron star occurs.   We then  consider aspects of superfluidity in dense matter.  Estimates of pairing gaps in homogeneous nuclear matter are given, and the effect of the dense medium on the interaction between nucleons is described.  Finally, we turn to superfluidity in the crust of neutron stars and especially the neutron superfluid density, an important quantity in the theory of sudden speedups of the rotation rate of some pulsars.  
\end{abstract}
\PACS{21.65.Cd  21.65.Mn 97.60.Jd}
  
\section{Introduction}
The purpose of this series of lectures is to describe in simple terms a number of properties of dense matter of importance for neutron stars.  In the main, the emphasis is on physical principles and obtaining a good semiquantitative understanding without going into fine details.  We shall adopt the format of a series of   ``Frequently Asked Questions'' and we begin in Sec.~2 with an overview of matter at subnuclear densities.  Section \ref{pasta} is devoted to phases in which nuclei can adopt rod-like and plate-like forms, as opposed to the roughly spherical shapes familiar terrestrially, and Sec.~\ref{superfluidity} considers aspects of neutron superfluidity and proton superconductivity in neutron stars.  We shall restrict ourselves to matter at subnuclear densities because, while such matter makes up only a small fraction of the total mass of a neutron star, many observational effects depend crucially on the properties of the outer layers of the star.  Matter at higher densities is reviewed in Ref.\ \cite{Baym}. 

In most of the discussion, we shall consider the ground state of matter as a function of the density of baryons.  Effects of nonzero temperature on the equation of state of matter are generally small, since in neutron stars the thermal energy $k_BT$ is small compared with characteristic microscopic energies (typically on the scale of MeV, or $10^{10}$K) except close to the surface of the star.  Of course, the temperature is important for thermal effects such as the heat capacity and nonequilibrium phenomena such as transport coefficients and neutrino emission rates. 

\section{Matter at subnuclear densities}
\label{faq}
Determining the properties of matter at terrestrial densities is a complicated problem because it demands a detailed understanding of electron correlations.  At higher densities, interactions between electrons become less important and the problem becomes simpler \cite{PethickRavenhall}.  
\subsection{How important are electron--electron interactions?}
\noindent A dimensionless measure of the importance of electron--electron interactions is $e^2/\hbar v_e$, where $v_e$ is the electron Fermi velocity.

{\centering \ldots  \par}
The density of a uniform electron gas is given in terms of the electron Fermi momentum $p_e$ by 
\be
n_e=\frac{p_e^3}{3\pi^2 \hbar^3}.
\ee
If one imagines the volume per electron to be a sphere,  its radius  is given by
\be
n_e  \frac{4\pi r_e^3}{3}=1\,\,\,{\rm or}\,\,\, r_e=\left(\frac{9\pi}{4}  \right)^{1/3}\frac{\hbar}{p_e}.
\label{re}
\ee
The Coulomb interaction energy of two electrons separated by a distance $r_e$ is $e^2/r_e$, while the kinetic energy of an electron is of order the Fermi energy, which is $p_e^2/(2m_e)$ for nonrelativistic electrons, where $m_e$ is the electron mass.  So as not to obscure the basic physics, we shall frequently omit numerical factors.    For an ultrarelativistic electron, the Fermi energy is $p_e c$, where $c$ is the velocity of light, and therefore quite generally, the typical kinetic energy of an electron is of order $p_e v_e$.  Thus the ratio of a typical interaction energy compared with the kinetic energy is of order
\be
\frac{e^2}{r_e p_ev_e}  \simeq \frac{e^2}{\hbar v_e} ,
\ee
from which one concludes that electron--electron interactions become \textit{less} important as the density increases, in contrast to what is the case for other sorts of interactions, such as those between atoms and molecules or between nucleons in nuclei and nuclear matter.  For nonrelativistic electrons, this ratio may be rewritten as ${r_e}/{a_0}$ where $a_0=\hbar^2/m_ee^2$ is the Bohr radius.  The ratio  ${r_e}/{a_0}\equiv r_s$ is the standard measure of the interaction strength used in the theory of metals at terrestrial densities, for which $r_s$ is of order unity  \cite{Pines}.   

\subsection{How important are electron--nucleus interactions?}
\noindent A dimensionless measure of the importance of electron--nucleus interactions on the properties of the electrons is $Z^{2/3}e^2/\hbar v_e$, where $Z$ is the atomic number of the nucleus. 

{\centering \ldots  \par}
The magnitude of a typical energy of interaction between a nucleus and an electron is $E_{eN}\simeq Ze^2/r_c$, where $r_c$ is a typical distance of an electron from a nucleus, which,  by analogy with the definition of $r_e$, Eq.~(\ref{re}), we define by the equation
\be
n_N\frac{ 4\pi r_c^3}{3}=1,
\label{rN}
\ee 
where $n_N$ is the density of nuclei.
Since bulk matter is electrically neutral, the densities of electrons and protons are the same, and therefore $n_e=Zn_N$, or
$r_c=Z^{1/3} r_e$.  Thus we see that the ratio of the Coulomb energy to the kinetic energy is of order $Z^{2/3}e^2/\hbar v_e$.  Thus electrons are little affected by interactions with nuclei if $Z^{2/3}e^2/\hbar v_e \ll 1$: matter is said to be ``pressure ionized''.

Another way of deriving this condition is by considering when  electrons can move relatively easily from one atom to another.  In Thomas--Fermi theory, the size of an atom is of order $a_0/Z^{1/3}$.  Atomic electron clouds will overlap when  the spacing between nuclei is less than this, or
\be
r_c\lesssim \frac{a_0}{Z^{1/3}}.
\ee 
With numerical factors inserted, interactions between electrons and nuclei will be small if the mass density, $\rho \gtrsim 10~AZ$ gm/cm$^3$, where $A$ is the mass number of the nucleus, the total number of nucleons per nucleus. 

Under terrestrial conditions, atoms behave as though their size is $\sim a_0$ rather than the Thomas--Fermi length $a_0/Z^{1/3}$.  This is because at low pressure properties such as the equilibrium density of matter are determined by the outermost electrons, which have a scale $\sim a_0$, and the majority of electrons in the core of the atom play little role.

\subsection{When do electrons become relativistic?}
\noindent At a density of order $10^6$ gm/cm$^3$.

{\centering \ldots  \par}
For electrons to be relativistic, the Fermi momentum must be comparable to $m_e c$.   In matter at ordinary densities, $\sim 1$ gm/cm$^3$, the separation between electrons is $\sim a_0$. A typical electron momentum is $\sim \hbar/a_0$ and a typical velocity $\hbar/(m_e a_0)=\alpha c$,  where $c$ is the speed of light and $\alpha=e^2/(\hbar c)$ is the fine structure constant.  The number $10^6$ can be understood simply since if one decreases the separation between electrons by a factor $\alpha$, thereby increasing the density by a factor $1/\alpha^3\sim 10^6$, the electron momentum is of order $m_ec$.

\subsection{Why do nuclei become more neutron rich with increasing density?}

\noindent The increase of the electron Fermi energy with density makes it energetically favorable to decrease the number of electrons relative to nucleons.

{\centering \ldots  \par}
Consider the energy per unit volume of matter as a function of the density of neutrons, $n_n$, the density of protons, $n_p$, and the density of electrons.  For electrically neutral matter, the density of electrons is equal to that of protons.   Weak interactions that can convert neutrons to protons or vice versa by processes such as 
\be
n\rightarrow p+e + \bar{\nu}_e \;\;\;{\rm and}\;\;\;  p+e \rightarrow n +\nu_e.
\ee
If one neglects neutrino rest masses and treats the energy as a continuous function of the particle densities, the condition for neither process to occur is that 
the energy to add a neutron to the system should be the same as that to add a proton and an electron:
\be
\mu_n=\mu_p+\mu_e,
\ee  
where $\mu_i=\partial E/\partial n_i$ is the chemical potential of species $i$ including contributions from rest masses.  This is the condition for matter to be in equilibrium with respect to the weak interactions. The nuclei encountered in neutron stars have a significant number of nucleons, larger than that of $^{56}$Fe, the most stable nucleus at low electron densities, and it is therefore a good approximation to use the liquid drop model to describe nuclear masses.  In addition, surface and Coulomb energies are small compared with bulk energies, so as a first approximation it is sufficient to take into account only bulk energies and write the contribution to the energy per unit volume from nucleons as
\be
E_N=n_nm_nc^2+n_pm_pc^2+n(-b+S \delta^2)
\label{energydensity}
\ee  
and neglect surface and Coulomb energies.
Here $n=n_n+n_p$ is the total nucleon density, $b$ is the bulk binding energy per nucleon in nuclear matter with equal numbers of neutrons and protons, empirically $\approx 16$ MeV, $S$ is the so-called symmetry energy,  empirically $\approx 32$ MeV, and 
\be
\delta =\frac{n_n-n_p}{n_n+n_p}=1-2x
\ee
is the neutron excess. Here $x=n_p/n$ is the proton fraction.  The neutron excess is zero for symmetric nuclear matter and unity for pure neutrons.

From expression (\ref{energydensity}) one finds that
\be
\mu_n-\mu_p=  (m_n-m_p)c^2+4S\delta.
\ee
When the electrons are relativistic, $\mu_e=p_ec$, and the electron density is given by
 \be
 n_e=\frac{1}{3\pi^2}\left(\frac{\mu_e}{\hbar c}\right)^3 \approx \frac{1}{3\pi^2}\left(\frac{(m_n-m_p)c^2+4S\delta}{\hbar c}\right)^3,
 \ee
which shows that the neutron excess increases with electron density.
\subsection{Why do neutrons drip out of nuclei at a density much below nuclear density?}
\noindent Because the electron velocity is considerably higher than the velocity of nucleons in nuclei, and because the binding energy of symmetric nuclear matter is considerably less than the Fermi energy of a nucleon.  

{\centering \ldots  \par}
The condition for neutrons to drip out of nuclei is that the neutron chemical potential exceed the neutron rest mass, in which case neutrons can propagate in the space between nuclei.  From Eq.~(\ref{energydensity}), one finds for the neutron chemical potential the expression
\be
\mu_n-m_nc^2\approx -b+S\delta(2-\delta).
\label{mun}
\ee
If one takes only the linear term in $\delta$ in this equation, one finds
\be
\delta_{\rm drip}\approx \frac{b}{2S}\approx \frac14,
\ee
and the electron density is 
\be 
n_e\approx \frac{1}{3\pi^2}\left(\frac{2b}{\hbar c}\right)^3.
\ee
The nucleon density at neutron drip is thus
\be
n_{\rm drip} \approx \frac{1}{3\pi^2 x_{\rm drip}}\left(\frac{2b}{\hbar c}\right)^3.
\label{ndrip}
\ee
As one can see, the $c$ occurs here because it is the electron velocity. It is helpful to introduce the Fermi momentum $p_s$, the Fermi velocity $v_s$ and the 
Fermi energy of a nucleon, $E_{s}=p_s^2/(2 m_n)$ at the zero-pressure (saturation) density of matter with equal numbers of neutrons and protons, $n_s\approx 0.16$ fm$^{-3}$.  The latter is given by  
\be
n_s=\frac{2}{3\pi^2}\left(\frac{p_s}{\hbar}\right)^3.
\ee
Equation (\ref{ndrip}) may then be written in the form
\be
n_{\rm drip}\approx \frac{1}{2x_{\rm drip}}\left(\frac{b}{E_s}\frac{v_s}{c}\right)^3 n_s.
\label{ndrip1}
\ee
This shows that, since the binding energy of nuclear matter is less than one half of the Fermi energy ($\approx 37$ MeV) and the nucleon velocity $v_s$ is about one third of the electron velocity, the density at neutron drip is more than two orders of magnitude less than that of saturated nuclear matter.

If the quadratic term in $\delta$ in Eq. (\ref{mun}) is included, one finds $x_{\rm drip}\approx 0.29$ with the values for the bulk and surface energies we have used.  This is close to what one finds from more detailed calculations \cite{CJPRavenhall}.  That the agreement is so good is fortuitous because of a cancellation between surface and Coulomb energies and the effect of nucleon shell structure, which is not included in the liquid drop model.

\subsection{At what density do nuclei disappear?}
\noindent At a density roughly the density of matter in nuclei.

{\centering \ldots  \par}
When the total density of matter is above the density of matter in nuclei, nuclear matter will occupy the whole of space, and therefore there will no longer be isolated nuclei. In Sec.~2 we shall consider the exotic shapes of nuclear matter at densities just below the saturation density.
\subsection{What are the atomic numbers of nuclei in the outer parts of neutron stars?}\label{nuclearsize}
\noindent They range from 26 (iron) at the surface of the star to around 40 in the inner crust where nuclei coexist with neutrons.

{\centering \ldots  \par}
If only bulk contributions are included in the nuclear energy, the energy per unit volume depends only on the total nucleon density and the proton fraction and is independent of the nuclear size. The equilibrium size of nuclei depends on contributions to the energy per nucleon beyond the bulk ones, the most important of which are the surface and Coulomb energies.  The former is proportional to the nuclear surface area, which scales as the square of the nuclear radius $r_N\propto A^{1/3}$ since the density of nucleons in nuclei is close to the saturation density, and one may write the  surface energy of a single nucleus as
\be
{\cal E}_{\rm surf}=C_{\rm surf}A^{2/3},
\ee
 where empirically the coefficient $C_{\rm surf}$ is $\approx 18$ MeV.  We use the symbol $\cal E$ to denote the energy per nucleus.  The Coulomb energy of a uniformly charged sphere of radius $r_N$ and total charge $Ze$ is $(3/5)Z^2e^2/r_N$.  For a nucleus, $r_N \propto A^{1/3}$ and therefore 
 \be
{\cal E}_{\rm Coul}=C_{\rm Coul}\frac{Z^2}{A^{1/3}},
\label{ECoul0}
\ee
 where, empirically, $C_{\rm Coul}\approx 0.7$ MeV.
To determine the optimal nuclear size, let us imagine that the proton fraction of matter, $x=Z/A$ is held fixed and ask for what $A$ the energy per nucleon is a minimum.  Since ${\cal E}_{\rm surf}/A\propto A^{-1/3}$ and ${\cal E}_{\rm Coul}/A \propto Z^2/A^{4/3}=x^2A^{2/3}$, it follows that the minimum energy is achieved for 
\be
{\cal E}_{\rm surf}=2{\cal E}_{\rm Coul},
\label{Esurf=2ECoul}
\ee
or 
\be
A\approx \frac{13}{x^2}.
\ee
This is consistent with the fact that the nuclei having the largest binding energy per nucleon with roughly equal numbers of neutrons and protons have $A\sim 50$.
At low densities, the favored nucleus is in fact $^{56}$Fe if one uses experimentally determined nuclear masses, but with increasing density the proton fraction falls, which leads to an increase in $A$.  

Two other effects have to be taken into account.  One is that, with decreasing proton fraction, the surface tension, the energy per unit area of the surface, is reduced, which tends to decrease $A$.  A second effect is that,  when the nuclear radius is no longer small compared with the separation between nuclei, the total Coulomb energy is reduced due to contributions from nucleus--nucleus, nucleus--electron and electron--electron interactions. The total Coulomb energy of a nucleus and the neutralizing cloud of $Z$ electrons, which are taken to lie within a sphere of radius $r_c$, is
\be
{\cal E}_{\rm Coul}= \frac{3}{5}\frac{Z^2e^2}{r_N}\left(1-\frac{3}{2}\frac{r_N}{r_c} +\frac12 \left( \frac{r_N}{r_c}\right)^3 \right).
\label{ECoulMatter}
\ee
Thus the total Coulomb energy is reduced as the density increases.  For example, at a density three orders of magnitude below nuclear matter density,  which is roughly the case for neutron drip to set in, $r_N/r_c=1/10$, and the Coulomb energy is reduced by 15\%. When nuclei fill all of space, the Coulomb energy vanishes because the proton and densities are equal everywhere, not just on average.

The condition for the equilibrium nucleus, Eq.~(\ref{Esurf=2ECoul}), still holds when the additional contributions to the Coulomb energy are included, the surface energy coefficient depends on proton fraction, and there are neutrons between nuclei, as is the case at densities above that for neutron drip.  The surface energy of nuclear matter is reduced dramatically for small proton concentrations, and for $x=0.1$ it is more than an order of magnitude less than for symmetric nuclear matter.  When these effects are taken into account, the atomic number remains around $40$ throughout the crust at densities higher than that for neutron drip.

 \section{Pasta phases}
 \label{pasta}
 At densities approaching that nuclear saturation density, nuclei can adopt forms very different from the roughly spherical nuclei familiar on Earth.  Because of the rod-like and plate-like forms of nuclear matter that result, these states have been dubbed ``pasta phases" because of their resemblance to spaghetti and lasagna. When the volume fraction occupied by pure neutron matter becomes less than one half, it is energetically favorable for nuclei to turn inside-out, but we shall not dwell on these phases because detailed calculations indicate that they are less prevalent. 
\subsection{Why do the pasta phases arise?}
\noindent Because of the competition between Coulomb and surface energies.

{\centering \ldots  \par} 
We have seen in Sec.~\ref{nuclearsize} that the equilibrium size of nuclei at low densities is determined by competition between surface and Coulomb energies.  
Similar arguments apply for rod-like and plate-like nuclei.  We shall characterize the nuclear shapes by their dimensionality  $d=$ 3, 2, and 1, corresponding to spherical nuclei, spaghetti and lasagna, respectively.  We shall denote the volume fraction occupied by nuclear matter by $u$,  and the radius of a spherical nucleus, the radius of a spaghetti strand, or half the thickness of a sheet of lasagna by $r_N$.  The surface energy per unit volume is given by \cite{pasta} \cite[Sec.~5.1]{CJPRavenhall}
\be
E_{\rm surf}=\frac{du\sigma }{r_N}
\label{Esurfd}
\ee
and the Coulomb energy by
\be
E_{\rm Coul}=2\pi(n_i x_i e )^2 u f_d(u)r_N^2, 
\label{ECould}
\ee
where 
\be
f_d(u)=\frac{1}{d+2}\left[ \frac{2}{d-2}\left(1-\frac{du^{1-2/d}}{2}\right) +u   \right].
\ee
For small filling factors $u$, $f_3$ tends to a constant, while $f_2$ diverges as $\ln u$ and $f_1$ diverges as $1/u$.  The large Coulomb energies at small $u$ disfavor nonspherical nuclei at low density but other shapes become favorable at higher values of $u$.  

From the fact that the surface and Coulomb energies scale with $r_N$ in the same way for the various shapes of nuclei, it follows that the equilibrium condition is
\be
E_{\rm surf}=2E_{\rm Coul}.
\label{Esurf2ECoul}
\ee
\subsection{Is there a simple physical picture that can help explain the formation of pasta phases?}
\noindent Yes, the fission instability for a charged liquid drop.  However, there are complications.

{\centering \ldots  \par} 
As shown by Bohr and Wheeler in their famous paper on nuclear fission \cite{BohrWheeler},  an isolated  charged, initially spherical liquid drop is unstable to a small quadrupolar distortion of its surface if 
\be
{\cal E}^0_{\rm Coul} > 2{\cal E}_{\rm surf},
\label{fission}
\ee
where the superscript ``0'' denotes the fact that the Coulomb energy is to be evaluated for low matter densities, (see Eq.~(\ref{ECoul0}).  Thus we see that for the equilibrium isolated nucleus at low density, Eq.~(\ref{Esurf=2ECoul}), the Coulomb energy is a factor 4 too small to cause fission, which is consistent with iron with $A=56$ being the most stable nucleus, while fission sets in only for $A \gtrsim 240$, e.g., uranium.  

In dense matter, the result (\ref{ECoulMatter}) indicates that the Coulomb energy is reduced by a factor $\simeq 1-3r_N/r_c$ for small $r_N/r_c$.  However, corrections to the fission condition due to surrounding matter may be shown to be of higher order \cite{Brandt}.  Thus one concludes that fission of spherical nuclei would set in when $1-3r_N/r_c \approx 1/4$, or $r_N\approx r_c/2$, which corresponds to nuclei filling one eighth of space.  One can imagine round nuclei becoming elongated and joining to make elongated structures like spaghetti.   While this picture is suggestive, it needs refinement.  Within the liquid drop picture, distorting a collection of spheres to make a collection of rods involves a change in the topology of the nuclear surface.  Thus on general grounds,  one would expect the transition to be of first order.

The problem of determining the equilibrium structure of a charged fluid of constant density with a uniform charged background to ensure electrical neutrality of bulk matter is a typical example of frustration: the surface energy is minimized by making the surface area as small as possible, thereby favoring structures with long length scales, while the Coulomb energy is minimized by breaking up the charged fluid into small drops.  Recently, Kubis and W\'ojcik have investigated the stability of the phase  with uniform lasagna sheets to arbitrary small distortions of the surface  \cite{KubisWojcik}.  They find that the phase is always stable to second order in the displacements from the uniform state.  Further inspiration can be derived from quantum molecular dynamics simulations of collections of neutrons and protons \cite{Watanabe}.  In these simulations, the particles are treated as classical, and effects of the Pauli exclusion principle are mocked up by introduction of an additional repulsive two-body interaction.   The simulations were designed to investigate whether or not pasta structures could form in the collapse of a massive star, and it was found that, on compressing a lattice of $^{208}$Pb nuclei, neighboring pairs of nuclei approached each other, and joined to form a zig-zag structure which subsequently became an array of rod-like nuclei.  It is important to bear in mind that these simulations were performed with a proton fraction, $x=0.39$, higher than that expected in neutron stars.   There are clearly many unanswered questions related to the pasta phases. 

So far we have focused on the problem of determining the optimal shape of nuclear matter when there is a separation into one phase with nuclear matter containing protons and one with pure neutrons.   It is also important to investigate whether the energy of the pasta phase is less than that of a uniform liquid of neutrons and protons.  For the FPS nucleon--nucleon interaction, which had been carefully fitted to calculated properties of pure neutron matter, this was indeed the case \cite{LorenzRavenhallCJP}.   Bao and Shen have carried out a parameter study of a set of relativistic mean field theory interactions that were fitted to the properties of symmetric nuclear matter \cite{BaoShen}.  They find that the existence of the pasta phases appears to be correlated with the parameter $L$, the derivative of the symmetry energy with respect to the logarithm of the total nucleon density,  and that the pasta phases are never the most stable state for $L\gtrsim 80$ MeV.     In the light of this result, it would be interesting to explore the reason for the interaction used in Ref.~\cite{DouchinHaenselMeyer} not giving pasta phases as the most stable ones.   Currently, experiment and theory favor values of $L$ lying between approximately 35 and 65 MeV \cite{HebelerSchwenk}, and the most recent chiral effective field theory calculations give values lying in the upper part of this range, 58.3--68.5 MeV \cite{DrischlerHebelerSchwenk}.   The quest to understand the pasta phases thus provides an incentive for studying in greater detail the microscopic nucleon--nucleon interactions used in nuclear physics, a task which is also important for improving interactions to be applied to neutron-rich nuclei in the laboratory.   

The properties of the inner part of the crust of neutron stars have implications for observation.  Quantities of interest include the elastic properties and breaking strain, which are important for predicting continuous gravitational waves from rotating neutron stars and for stellar oscillations.  

\subsection{What are the elastic properties of the pasta phases?}
\noindent They are similar to those of liquid crystals, since some distortions, like sliding lasagna plates over each other, give no restoring forces.

{\centering \ldots  \par} 
As an example, let us consider the lasagna phase, with the plates lying in the $x$-$y$ plane.  A distorted configuration may be described by the displacement of a plate in the $z$-direction, $u_z$ and the only contribution to the elastic energy per unit volume to lowest order in spatial gradients is 
\be
E_{\rm elast}=\frac12 c_{33}\left(\frac{\partial u_z}{\partial z}    \right)^2,
\ee
where $c_{33}$ is the corresponding elastic constant.\footnote{In order to make the treatment accessible to a broader readership, we here use the Voigt notation standard in elastic theory rather than the notation used in the literature on liquid crystals \cite{deGennesProst} and the pasta phases \cite{CJPPotekhin}.}  We shall here consider distortions that are not accompanied by changes in the mean densities of protons and neutrons.  The bulk energy densities  and the fraction of space filled by nuclear matter remain unchanged and the only the thickness of the lasagna sheets and the spacing between them change.  When $r_N$ is increased by an amount $\delta r_N$, the strain ${\partial u_z}/{\partial z} $ is $\delta r_N/r_N$.  Since the surface energy per unit volume (\ref{Esurfd}) scales as   $r_N^{-1}$ and the Coulomb energy (\ref{ECould}) as $r_N^2$, the second derivatives of these energies with respect to $r_N$ are given by $2(E_{\rm surf}+E_{\rm Coul})/r_N^2$, from which it follows that \cite{CJPPotekhin}
\be
c_{33}=2(E_{\rm surf}+E_{\rm Coul})=3E_{\rm surf}=   \frac{3\sigma}{r_c}, 
\ee
where $r_c$ for lasagna is one half the layer spacing, and we have used the equilibrium condition Eq.~(\ref{Esurf2ECoul}).  The spaghetti phase can be treated in a similar fashion but it is more complicated, since it has two second-order elastic constants, corresponding to homologous compression and shearing of the triangular lattice on which the centers of the spaghetti strands are located. 

\subsection{Are pasta elements uniform?}
\noindent Probably not.  Thomas--Fermi,  Hartree--Fock and molecular dynamics calculations indicate that the thickness of lasagna plates has periodic modulations in the plane of the plates, and that the thickness of spaghetti strands is modulated along the length of a strand. 

{\centering \ldots  \par} 
The numerical evidence for spatial modulation  of the spaghetti and lasagna phases is suggestive but, so far, there is little analytical work on the subject.  The  Thomas--Fermi \cite{WilliamsKoonin},  and Hartree--Fock \cite{PaisandOthers} calculations consider relatively small cubic, cells of matter, and it is unclear to what extent the results are influenced by the boundary conditions.  The molecular dynamics calculations \cite{Berry, Caplan} use much larger cells, but here the question is whether the method, which is essentially classical, is able to capture the properties of cold, dense systems of nucleons. 
Spatial modulation of the pasta structures has important consequences for the properties of these phases.  The low-frequency collective modes of the structure would resemble those of an anisotropic three-dimensional solid, rather than a liquid crystal \cite{KobyakovCJP2020}.  The work of Peierls and Landau \cite{Peierls, Landau} shows that one-dimensional ordering of the density will be destroyed by thermal fluctuations.   This result is no longer applicable if there are modulations in three dimensions.

\section{Superfluidity}
\label{superfluidity}

\subsection{Why are neutron superfluid gaps in neutron matter so small?}
\noindent While the interaction between two neutrons in different spin states is attractive at low densities, at even small neutron densities the effect of the repulsive part of the interaction becomes very significant.

{\centering \ldots  \par} 
In neutron matter, pairing gaps are estimated to have values of at most  $\sim 1$ MeV, while the Fermi energy at a density of half nuclear matter density is $\sim$ 37 MeV.  Properties of cold Fermi gases with strongly attractive interactions have been studied intensively over the past two decades and superfluid gaps can be a significant fraction of the Fermi energy.  This is also the case for neutrons, but only at very low densities.  For pairing, the relative momenta of importance for  the superfluid gap are of order the Fermi momentum, $p_n$.  While for neutrons the interaction is strongly attractive at very low relative momenta, the effect of the repulsive short-range part of the interaction becomes important for relative momenta much less than the Fermi momentum at nuclear matter density, $p_s /\hbar \sim 1$ fm$^{-1}$. 

Consider pure neutron matter.  The interaction between two neutrons in opposite spin states is attractive at low relative momenta and is almost strong enough to make a bound state of two neutrons, the dineutron.  Quantitatively, the effective interaction between neutrons in vacuo at low relative momenta is given in terms of the scattering length, $a$, by
\be
U_0=\frac{4\pi \hbar^2a}{m_n}.
\label{contact}
\ee
If the pairing interaction is given by this expression and the effect of the neutron medium on the pairing interaction is neglected, one finds for the neutron superfluid gap at low densities \cite{GorkovMB}
\be
\Delta=\frac{8}{\rme^2}E_n\rme^{-\pi/(2k_n|a|)},
\label{gapBCS}
\ee
for negative $a$ (attractive interactions).  Here $E_n$ is the neutron Fermi energy.  For higher densities, $k_n |a|\gtrsim 1$, the gap is comparable to the Fermi energy and the state resembles a Bose--Einstein condensate of diatomic molecules. The experimental value of $a$ is $-18.5$ fm, which is much greater than the range of the neutron--neutron interaction, $\sim 1$ fm.  Formation of a bound state is signaled by the condition $a\rightarrow \infty$.    Thus for low densities, one expects the gap to rise with density according to Eq.~(\ref{gap}) until it reaches a value comparable to the Fermi energy for $k_n|a| \sim 1$, or a density $\sim |a|^{-3}\approx 10^{-3}$ fm$^{-3}$, which is much less than the density of matter in nuclei \cite{warning}.

In the remainder of this subsection we shall consider the interaction between neutrons to be given by its value in free space, which we shall refer to as the BCS (Bardeen--Cooper--Schrieffer) approximation.  The S-wave interaction between two particles is characterized by the phase shift, $\delta_0(k)$, where $\hbar k$ is the relative momentum.  For short-range interactions and for small $k$, the phase shift may be written in the form
\be
\frac{1}{k\tan \delta_0}\simeq -\frac{1}{a} +\frac12 r_{\rm eff}k^2,
\ee
 where $r_{\rm eff}$ is the effective range. The  magnitude  of $r_{\rm eff}$ specifies how the importance of the repulsive part of the interaction grows with $k$.  For neutrons, both the scattering length, $\approx -18.5$ fm, and the effective range, $\approx 2.7$ fm, are large in magnitude compared with the range of the neutron--neutron interaction, $\sim 1$ fm.  Thus one sees that the effective range term is significant for 
 \be
 k_n\gtrsim k_\times = \frac{2}{(|ar_{\rm eff}|)^{1/2}}\approx 0.28~{\rm fm}^{-1}.
 \ee
A neutron gas with a Fermi momentum equal to $k_{\times}$ has a density roughly 1 \% of nuclear matter density.  

The overall picture that emerges is that the superfluid gap rises with the neutron Fermi wave number according to Eq.~(\ref{gapBCS})  and reaches a value comparable to the Fermi energy for $k_n\sim |a|^{-1}$.  For larger wave numbers it remains comparable to the Fermi energy until $k_n \sim (|a| r_{\rm eff})^{-3/2}$, after which the repulsive part of the interaction plays an important role and suppresses the gap. This is illustrated in Fig.~1, where $\Delta/E_F$ is plotted against the Fermi wave number for the contact interaction (\ref{contact}) appropriate for cold atoms and for a model of the neutron--neutron interaction. To calculate the gap at higher densities demands use of more realistic expressions for the interaction.  Calculations that assume the pairing interaction to be equal to that between two neutrons in vacuo give S-wave gaps that have a maximum of around 3 MeV at a density of about one eighth of nuclear density, and  vanish for  densities greater than 0.6 $n_s$.

\begin{figure}
\includegraphics[width=3.5in]{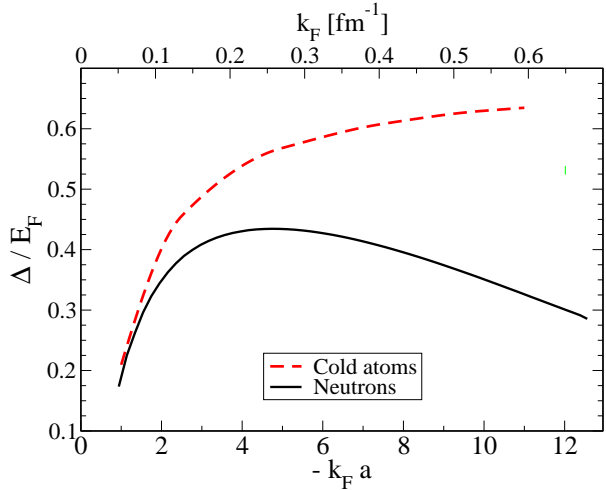}
\caption{Ratio of the pairing gap in the BCS approximation to the Fermi energy as a function of Fermi wavenumber for two different potentials, a short-range one, Eq.~(\ref{contact}) appropriate for cold atomic gases, and a model of the neutron--neutron interaction.  While $\Delta/E_F$ continues to rise with increasing $k_F$ for cold atoms, for neutrons it reaches a maximum and subsequently falls because of the increasing importance of the repulsive part of the interaction.  The upper scale gives the neutron Fermi wave number when $a$ is taken to be the experimentally determined value $-18.5$ fm. For neutrons $\hbar k_F$ is equal to the neutron Fermi momentum, which is denoted by $p_n$ in the text. (Figure courtesy of A.~Gezerlis \cite{GezerlisCarlson}.  Reproduced from Ref.~\cite{GezerlisCJPSchwenk} with permission from Oxford University Press.)}
\label{Fig_1}
\end{figure}

\subsection{How does the neutron medium affect neutron superfluid gaps?}
\noindent  Exchange of spin fluctuations in the neutron medium reduce the superfluid gap to values having a maximum in the range 1-1.5 MeV.

{\centering \ldots  \par}
The interaction between two neutrons depends on the direct interaction of the two particles themselves but also on effects induced by the presence of other neutrons.  Such induced interactions are familiar in metallic superconductors, where the exchange of density fluctuations (lattice phonons) is the origin of the attraction between electrons.  In neutron matter there are both density fluctuations  and spin fluctuations.  Exchange of density fluctuations tends to increase the gap, while exchange of spin fluctuations reduces it.    The suppression of pairing gaps is well known in metallic superconductors with a large magnetic susceptibility, e.g., palladium, where there are low-lying electron spin excitations \cite{BerkSchrieffer}.  In neutron matter and in ultracold Fermi gases, the effect of spin fluctuations dominates because there are three magnetic substates, corresponding to magnetic quantum numbers $m_S=0, \pm1$, as opposed to the single state for the density fluctuation, which carries zero spin. Surprisingly, these effects are important even for a dilute Fermi gas, and the gap is given by \cite{GorkovMB}
\be
\Delta=\left(\frac{2}{\rme}\right)^{7/3}E_n\rme^{-\pi/(2k_n|a|)},
\label{gap}
\ee
which is a factor $(4\rme)^{-1/3} \approx 0.45$ times the result (\ref{gapBCS}) with medium effects on the interaction omitted \cite{Heiselbergetal}.  Detailed calculations for neutron matter of the reduction of the gap due to the effects of the medium are of a comparable magnitude.

The calculation of S-wave gaps in neutron matter is under good control at low densities, but the uncertainties become greater with increasing density.  There are two basic reasons: the gap depends exponentially on the pairing interaction, and the effect of the medium on the pairing interaction becomes increasingly important.  There is still significant uncertainty in the density at which S-wave superfluidity of neutrons ceases, at around a half of nuclear matter density. 

At higher densities,  the neutron--neutron interaction is most strongly attractive in the $^3$P$_2$ state, which is coupled by the tensor force to the $^3$F$_2$ state.  While S-wave gaps depend on the average of the pairing interaction over directions on the Fermi surface, those for other states depend on the average of the pairing interaction weighted by a function whose average vanishes.  Thus gaps for states other than S-wave ones depend on \textit{deviations} of the pairing interaction from its average value.  In addition, at the densities where  pairing in the $^3$P$_2$ state could be relevant, medium effects are important.  Calculations of the $^3$P$_2$ gap including medium effects come to the conclusion that gaps are very small \cite{SchwenkFriman}, and they could even vanish.
\subsection{Are there observable consequences of neutron superfluidity?}
\noindent A leading model to account for sudden speed-ups (glitches) in the rotation rate of pulsars is a neutron superfluid weakly coupled to the lattice of nuclei in the crust.  A sudden unpinning of neutron vortices from the nuclei results in slowing down of the neutron superfluid and a sudden increase in the angular velocity of the crust.  

{\centering \ldots  \par}
As a neutron star loses rotational energy, the crust and the charged particles coupled to it slow down.  On the other hand, because of the superfluidity, the rotation rate of the neutron superfluid remains higher than that of the crust.  In a superfluid, rotational motion is achieved by having an array of quantized vortex lines, each with circulation $2\pi\hbar/(2m_n)$.   The basic idea of the model is that, initially, vortex lines are pinned to nuclei, but when the difference between the rotation rates of charged particle and the superfluid becomes sufficiently large, the vortex lines unpin from the nuclei, thereby leading to a slowing down of the superfluid and a speed-up of the crust, whose rotation rate determines the observed pulsar frequency \cite{AndersonItoh, HaskellMelatos}. For this model to be viable, the moment of inertia of the superfluid must be large enough to be able to account for the observed magnitudes of glitches, and how frequently they occur in a given pulsar.  

To determine the moment of inertia of the superfluid neutrons it is thus necessary to know the superfluid density of neutrons in the presence of the lattice of nuclei, either in the form of a body centered cubic (b.c.c.) lattice of spherical nuclei or the pasta phases.  Chamel calculated the  band structure for neutrons in a b.c.c.~lattice in the absence of pairing, and concluded that, for weak pairing, the superfluid density could be reduced by more than one order of magnitude compared with the density of neutrons between nuclei, which is what one might naively expect \cite{Chamel2012}.   These calculations are very demanding because there are very many neutrons in a unit cell and therefore many bands (up to $\sim 500$).  On the basis of these calculations Andersson et al.~\cite{Andersson}, Chamel \cite{Chamel2013} and Delsate et al.~\cite{Delsate} concluded that the neutron superfluid moment of inertia was too small for the glitch model to be viable.

The question that then arises is whether the superfluid density is reduced as drastically when pairing is taken into account.  In an S-wave fermionic superfluid, the elementary excitations for momenta close to the Fermi momentum have energies
\be
\epsilon_p=\pm\left( \xi_p^2+\Delta^2   \right)^{1/2}
\ee
where $\xi_p\simeq (p-p_n)v_n$.  In the presence of a scalar potential such as that due to the periodic lattice of nuclei,  the amplitude for scattering a positive energy excitation with momentum $\pp$ to another state of positive energy with momentum $\pp'$ is proportional to 
\be
M=u_\pp u_{\pp'}-v_\pp v_{\pp'},
\ee 
where the coherence factors are given by
\be
u_{\pp}^2=\frac12\left( 1+ \frac{\xi_\pp}{\epsilon_\pp}      \right)\;\;\;{\rm and}\;\;\;  v_{\pp}^2=\frac12\left( 1- \frac{\xi_\pp}{\epsilon_\pp}\right). 
\ee
Thus for $p=p'=p_n$ the positive energy excitations are superpositions of particles and holes with amplitudes of the same magnitude and the scattering amplitude vanishes.  Expressed in physical terms, the potential acting on a particle is equal in magnitude but of opposite sign to that acting on a hole, so the net matrix element vanishes.  Similar arguments apply for scattering between two negative energy states.  The amplitude for scattering from a positive energy state to a negative energy one or vice versa is nonzero for $p=p'=p_n$.  Scattering between two positive energy states can occur via a second-order process with an intermediate negative energy state.  However, the amplitude for this process is of order $V^2/\Delta$, where $V$ is the strength of the scattering potential and the energy denominator in the intermediate state is $\sim \Delta$.   Thus one expects scattering by the lattice to be suppressed if the pairing gap exceeds the strength of the lattice potential.

In neutron star crusts, the pairing gap is greater than the strength of the periodic potential for most reciprocal lattice vectors, so it is necessary to include the effects of pairing in calculations of the superfluid density.  In Ref.~\cite{WatanabeCJP} this was done within the Hartree--Fock--Bogoliubov model, a mean field approach that treats the effects of pairing and the periodic potential on an equal footing.  In the calculations, a single Fourier component of the periodic potential was considered as had earlier been done for cold atoms in a one-dimensional periodic potential \cite{Watanabe2008}.  With the assumption that each Fourier component contributes independently to the reduction of the superfluid density, it was concluded that the periodic lattice could reduce the superfluid density by perhaps some tens of per cent, but not by an order of magnitude.  Consequently, the vortex unpinning model for explaining glitches could not be ruled out on the basis of the neutron superfluid density being too small.

\subsection{Are the pasta phases good or bad electrical conductors?}
\noindent  They could be very good conductors because the protons  are superconducting and can support persistent electrical currents.

{\centering \ldots  \par}

Pons et al.~\cite{Pons} have proposed that the lack of isolated X-ray pulsars with long periods could be explained if the pasta phases are poor electrical conductors due to scattering of electrons by disorder in the  pasta structures.  This could lead to decay of the magnetic field.  However, one effect that works in the opposite direction is that the protons in these phases are expected to be superconducting and, consequently, magnetic fields can be anchored in persistent proton currents in the pasta phases  \cite{Kobyakov}. 

In the pasta phases there are a number of other problems that ripe for investigation.  Among these are the elastic and superfluid properties of disordered phases.  

\section{Concluding remarks}

The problems considered in this article illustrate how basic physical principles can give an understanding of a number of the properties of matter in the outer parts of neutron stars. 
One important lesson is that in dense matter, what we normally consider to be ``solid-state'' energies can be significant on the scale of nuclear energies.

 There are many topics that have not been considered.  Among them are the properties of matter that is not in its lowest energy state, such as is the case for a neutron star which is accreting matter.  In addition, we have not investigated phenomena at non-zero temperature or transport properties.

 I am grateful to the organizers of the Zakopane summer school for the invitation to participate.  I should also like to thank Matt Caplan, Alexandros Gezerlis, Dmitry Kobyakov, Sebastian Kubis, Achim Schwenk and Gentaro Watanabe for helpful comments.


\begin{thebibliography}{99}
\bibitem{warning} The reader is warned that ``factors of order unity'' can be considerable.  For example, the density of a degenerate neutron gas is $n_n=k_n^3/(6\pi^2)$, and $6\pi^2\approx 59$.  In order to illustrate the physics as clearly as possible, generally no attempt is made to include such numerical factors.
\end{thebibliography}
\end{document}